\documentclass{aa}  

\usepackage{graphicx}
\usepackage{txfonts}
\usepackage[]{hyperref}

\begin{document}

   \title{Turning Spherical Cows into Spherical Cheeses:}

   \subtitle{A Bold and Flavourful re-Interpretation of the Moon's spectrum}

   \author{V. Yariv\inst{1}
          \and
          M. Ravet\inst{1, 2}
          }

   \institute{IPAG, Université Grenoble-Alpes, CNRS, F-38000 Grenoble,              France
        \and
            Laboratoire J.-L.; Lagrange, Université Côte d'Azur,
            Observatoire de la Côte d'Azur, CNRS, 06304 Nice, France}

   \date{Received April 1st, 2024; accepted YYYY}

% \abstract{}{}{}{}{} 
% 5 {} token are mandatory
 
  \abstract
   {"Everyone knows the moon is made of cheese..." This line, famously uttered by Wallace to his canine sidekick Gromit in the 80s classic, may be one of the most cruelly underappreciated movie quotations of our time.}
   {Indeed, while most scientists today would simply reject Wallace's claim as preposterous, we aim to revisit his theory on the composition of our natural satellite, revealing that it may not be as implausible as the scientific consensus would have it.}
   {Through a revelatory novel analysis of existing data, we will show that very simple cheese-based models can provide a convincing explanation of the Lunar surface's spectral characteristics in the near-infrared. Using the tried and tested PLS (Partial Least Squares) method, we efficiently and reliably retrieve the concentrations of various cheese types in different locations of the Lunar surface.}
   {Our results bring to light a bold and flavourful prediction about the Moon's composition, which lays the groundwork for an important paradigm shift in planetary sciences. We urge the scientific community to take a serious notice of this piquant novel interpretation, and strongly consider it in their future models of planetary composition and formation in our solar system and beyond.}
   {}


   \maketitle
%
%-------------------------------------------------------------------

\section{Introduction}

Scientific consensus today contends that planetary bodies in our Solar System and beyond are highly complex objects. Indeed, the great majority of scientists tend to argue for adding ever increasing complexity into their models, promptly arguing that planets and satellites "are not spherical cows" (\href{https://youtu.be/VyxMZ2vS3dI?si=MVzyNUNlX_N3aXpO}{Lehmann et al. (2022)}). Yet, while constantly multiplying the number of parameters in simulations may be an effective way of obtaining funding for this "complex" science, we would argue that it is actually evidence of important hubris on behalf of the investigators involved (as already suggested by the recent work of \citet{woodrum}). 

In pursuit of a more parsimonious epistemology of science, we turn to our closest celestial neighbour, the Moon (\href{https://www.allocine.fr/film/fichefilm_gen_cfilm=136189.html}{Jones et al. (2010)}). Thanks to its close proximity to us, the Moon has been a target of much study by the scientific community. In light of these many observations and measurements, theories on the Lunar composition and origin have been some of the worse offenders of complex science. Take for instance the lunar composition: scientists today generally propose a lunar regolith consisting of fragments of igneous rocks, crystalline impact melt rocks, meteoritic metals, breccias produced by meteoritic impacts, glassy spherules of volcanic origin among many other things (\citet{taylor}; \citet{tian}; ...). On top of requiring many obscure and cryptic science words to describe, this complicated composition is necessarily associated to a complex formation history. Some scientists have even gotten to the point of invoking a theoretical planetary body "Theia", which would have had to collide with the young Earth of all things to form our Moon (\citet{wood}; \citet{newsom}). Lunacy!

In this article, we return to basics and we hope to convincingly argue for a radically more simple and mature picture of the Moon: that it is made out of cheese. Through a novel data analysis of many previously published Lunar spectra, we will demonstrate how the conclusions heretofore wrought by the scientific community may be needlessly complicated. Spurred on by recent groundbreaking works in the field of spectroscopic characterisation of cheeses, we will show how a simple cheese-based model can effectively reproduce all of the Moon's notable spectral characteristics. Our model, combining just four of the most common cheese types (emmental, parmesan, cheddar and ricotta), provides a radical new paradigm on the nature of our celestial companion, and exemplifies how an exercise in humility about our prior suppositions can lead to groundbreaking science. In the end, we hope that this can be a lesson to our fellow scientists that by constantly building more complexity into a "spherical cow" model, we may be overlooking equally valid and interesting spherical cheese models, that would be worth digging into.

%--------------------------------------------------------------------

\section{Observation and data reduction}\label{sec1}

\begin{figure*}[ht]
    \centering
    \includegraphics[scale=0.17]{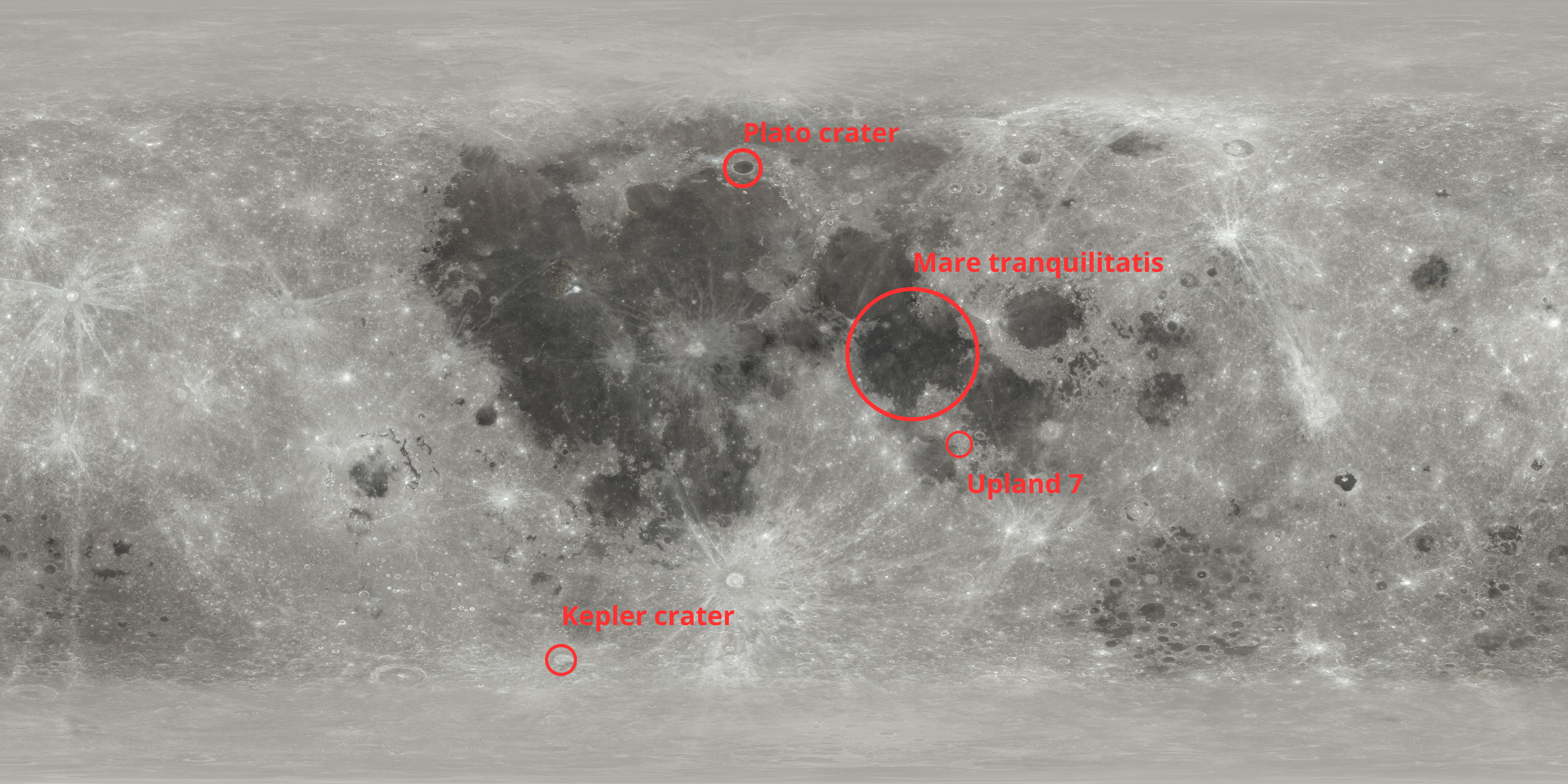}
    \caption{\small{Color-map of the lunar surface centered on $0^\circ$ latitude. Locations of the 4 observations taken from \citet{1981JGR....8610883M} are circled in red (\textit{image adapted from \href{https://svs.gsfc.nasa.gov/cgi-bin/details.cgi?aid=4720}{nasa-svs}}).}}
    \label{lunar_map}
\end{figure*} 

Given the proximity of the Moon to land and space based instruments, observational methods can provide spectra in both emission and reflection at very high spatial resolution. Collating a single comprehensive dataset of the Lunar spectrum can thus be a challenging task, and is beyond the scope of this paper. As a representative first sample, we chose to look at a set of spectra homogeneously compiled by \citet{1981JGR....8610883M}. These observations provide the advantage of being taken in the near-infrared (NIR), where the thermal emission contribution is negligible \citep{2017Icar..283..300W}, and can thus be readily compared to models.

The dataset from \citet{1981JGR....8610883M} compiles most of the NIR surveys of the lunar surface made by earth-based instruments. These spectra have been rigourously normalised in order to account for various biases (eg: instrumental calibration, ambient light conditions, human error or intentional meddling...), and we direct the reader to \citet{1981JGR....8610883M} for a more thorough description of these procedures. For our study, we focus on the spectra from the following four regions:

\begin{itemize}
    \item Plato crater
    \item Mare Tranquilitatis
    \item Upland 7
    \item Kepler crater
\end{itemize}

as they represent a diverse sample of the lunar surfaces (see figure \ref{lunar_map}). We estimate the spectral resolution at $R\sim100$, which is worse than 1000 but better than 10, and will do very well for efficient cheese comparison. Since \citet{1981JGR....8610883M} do not provide their dataset in an easily accessible format, we use state of the art human-assisted computational graphing techniques to recover the data directly from the plots in their article \footnote{This amazing tool is available for anyone to use at: https://apps.automeris.io/wpd/}. This revolutionary tool for enabling openness and reproducibility is unfortunately still rarely used in the community, but an upcoming article will hopefully change this in the near future.

We cut data between $1.350-2.350\,\mu m$, as this is the range in which we expect the most unique features in various cheese types, and where most cheese spectral studies overlap. We call this range the \textbf{REBLOCHON} (\textbf{RE}gion of \textbf{B}est \textbf{L}ines to \textbf{O}bserve \textbf{CH}eese \textbf{O}riginality in the \textbf{N}IR), for ease of referencing in the rest of the paper and future works. We additionally re-sample the observations into a wavelength grid consistent with our cheese models (see sec. \ref{sec2}) over the REBLOCHON range, in order to have a uniform analysis of all datasets.
%--------------------------------------------------------------------\ref{}

\section{Models}\label{sec2}

Our forward modelling approach is designed with simplicity and robustness in mind, in line with our guiding values of humble and open-minded science. Our principal goal, as stated in the introduction, is to propose a simple yet convincing cheese-based model of the lunar surface, which may then be built upon in collaboration with other like-minded scientists. As such, our analysis is built upon the \textbf{P}artial \textbf{L}east \textbf{S}quares (PLS) regression, which is a robust and efficient method to explore large parameter spaces containing multicollinearity. PLS makes use of a remapping of the parameter space in order to perform linear regression and obtain the best fitting parameter values. PLS then has two main advantages over other regression methods: firstly, it is linear, which as everyone knows means reliable and intelligible, and secondly, it is already implemented very thoroughly in python's \textsc{scipy} library, meaning that its implementation can be done in a mere couple lines of code, saving a great many time and efforts for the authors \footnote{If more funding were to be allocated to this project, we have plans to implement more complex regression methods such as the very fashionable MCMC (Mysteriously Computed Model Comparison) to extract posteriors with Bayesian statistics.}.

We introduce the \textbf{FROMAGE} (\textbf{F}lavorful and \textbf{RO}bust \textbf{M}oonlight \textbf{A}nalysis through \textbf{G}ourmet \textbf{E}missions) code, which can explore a wide range of bovine, caprine and even human cheeses (\href{https://youtu.be/uNlKjq3zl04?si=iIJ70x6M0cyZROI1}{Mamelou et al. (2016)}). Our FROMAGE code is incredibly user-friendly, consisting of entirely commented jupyter notebooks and containing various unique additional features, such as the possibility to patiently wait for the eponymous \textit{Godot}, an abbility which uses our bespoke implementation of the \textit{bogosort} algorithm \footnote{https://en.wikipedia.org/wiki/Bogosort}. The code is publicly available at the following \href{https://github.com/Qu4rkie/FROMAGE}{Github link}, along with all the datasets used in this work.

As our nominal starting model, we use an equal parts mixture of 4 common cheese types: Cheddar, Emmental , dry Parmesan and Ricotta (i.e. reference composition is 1/4 for each). The various reference cheese spectra being except from different publications using varied normalisation techniques (\citet{downey}; \citet{karoui}; \citet{frank} and \citet{madalozzo} respectively), we start by using visual inspection to re-normalise the models to a common scale such as not to bias the final measurement. Before inputting them into the FROMAGE code, we also convert the units from absorbance $A$ to reflectance $R$, using the conversion:
\begin{equation*}
   A=log(1/R) 
\end{equation*}

FROMAGE then takes as input the cheese spectra in reflectance, automatically cuts them over the REBLOCHON range and resamples them in a homogenous way with the Lunar data. Starting from the nominal model concentration, FROMAGE then uses Partial Least Squares to explore the ranges of possible concentrations for each cheese in the model, and find the best fitting one for each lunar region. The algorithm is quite efficient, and a good fit can be obtained within a few seconds for each region. As such, it is easily and feasibly possible to add complexity to the cheese models and consider more varied cheese types, which we intend to do in future work.

%--------------------------------------------------------------------

\section{Results and Discussion}\label{sec3}

\begin{figure}[h!]
\centering
    \includegraphics[scale=0.23]{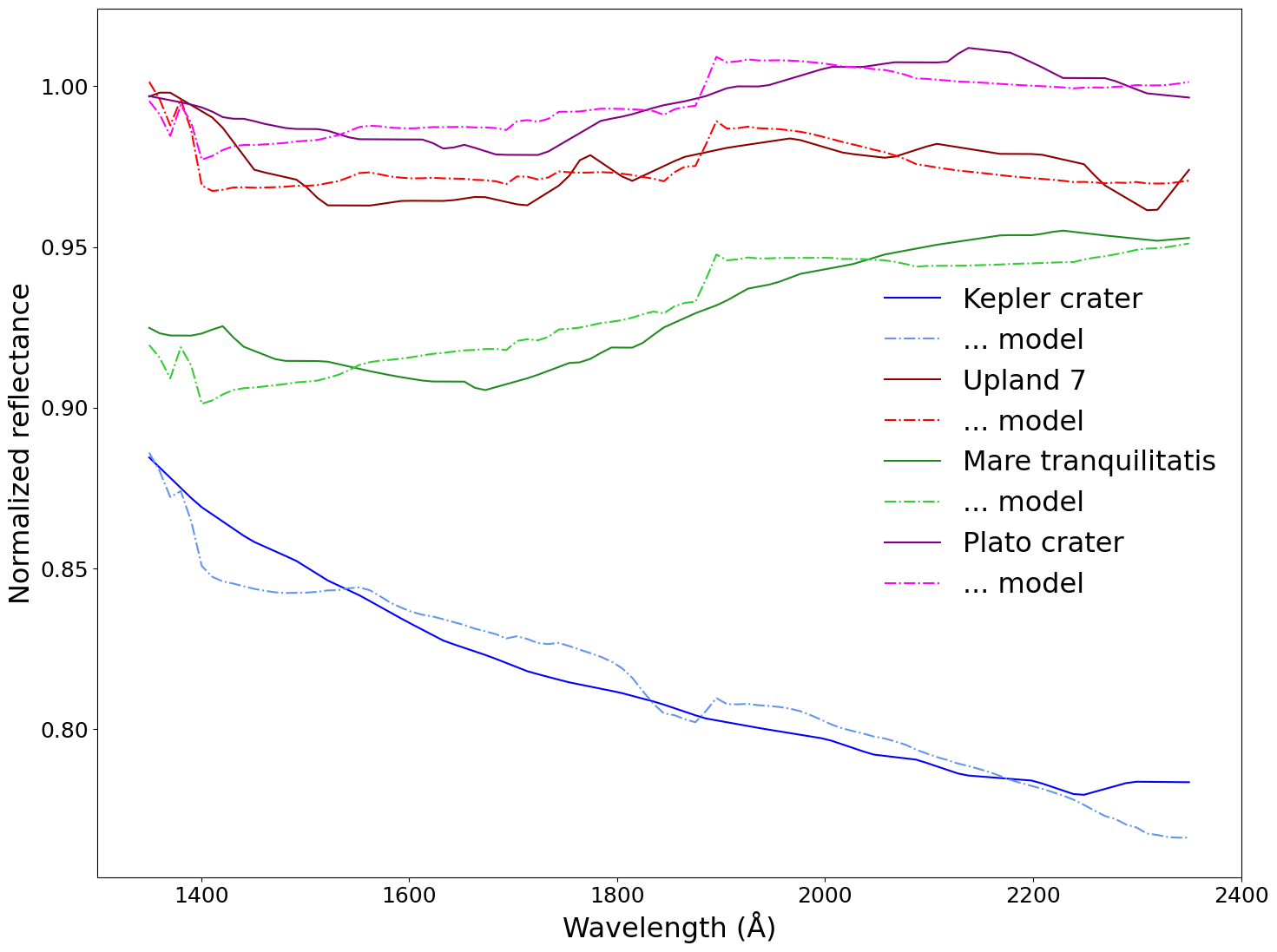}
    \caption{\small{Reflectance spectra (in solid lines) for each lunar regions (see section \ref{sec1}) compared with their best fitted FROMAGE models (in dotted lines). As can be seen, the match is very good between model and observation, and any deviations above the observations are subsequently compensated by deviation below the observations and vice-versa.}}
    \label{fit_moon}
\end{figure}

Results of our analysis on the composition of the lunar surface are presented in table \ref{posteriors}; the associated fitted reflectance spectra are plotted in figure \ref{fit_moon}. As expected, we find a high concentration of Emmental cheese over most Lunar regions considered, which explains the "cratered" aspect of the lunar surface, due to the infamous holes of the Emmental cheese. Equally, we find high concentrations of Cheddar cheese over all regions, which seems consistent with trends here on earth, where this cheese is one of the most widely exported and consumed throughout the planet.

\setlength{\tabcolsep}{6pt} % Default value: 6pt
\renewcommand{\arraystretch}{1.5} % Default value: 1
\begin{table}[h]
\tiny
    \centering
    \begin{tabular}{lllll}
    \hline
    \textbf{Cheese} & Emmental & Dry Parmesan & Cheddar & Ricotta \\
    \hline
    \textbf{Plato crater} & $34.2\%$ & $13.8\%$ & $39.3\%$ & $12.7\%$ \\
    \textbf{Mare Tranquilitatis} & $45.4\%$ & $3.1\%$ & $49.1\%$ & $2.4\%$ \\
    \textbf{Upland 7} & $48.5\%$ & $-1.6\%$ & $45.8\%$ & $7.3\%$ \\
    \textbf{Kepler crater} & $46.5\%$ & $2.0\%$ & $42.4\%$ & $9.1\%$ \\
    \hline
    \end{tabular}
    \\~\\
    \caption{\small{Best fitted cheese composition for each lunar region (see section \ref{sec1}). The statistical confidence interval at $99.8\%$ ($3\sigma$) is of $\pm 2\%$ for each measurment, as estimated by the FROMAGE code. Hence, we disregard the negative value found for Dry Parmesan concentration in Upland 7 as a statistical fluke, consistent with a $0\%$ concentration.}}
    \label{posteriors}
\end{table}

 Surprisingly, the Plato "cratered" region actually shows a lower concentration of Emmental cheese, but higher Ricotta and Dry Parmesan concentrations as compared to other regions. We interpret this as potential signs that said Ricotta and Parmesan have retroactively come and filled in the holes in the Emmental, which seems to be corroborated by the equally higher ricotta concentration in the Kepler region. In this case, the Parmesan concentration remains low however, which may be indicative of different epochs of Ricotta and Parmesan filling, with the Kepler hole being a more recent formation in the Emmental base layer. 

Finally, we note a very slightly negative concentration of Dry Parmesan in the Upland 7 region. While this may initially sound intriguing (my we have discovered anti-cheese?), a deeper analysis of the statistical error reveals that this is most likely a fluke, as the confidence interval at $3\sigma$ is actually consistent with this concentration being 0. This is quite reassuring, as it would be expected for cheese and anti-cheese to be extremely reactive together, and thus their simultaneous presence on the Lunar surface to be very unstable, leading to enormous energy discharges.

Despite these inconclusive results on the possibility of Lunar anti-cheese, the revelation of the cheese surface has important implications to simplify existing theories on the Moon's formation history. Indeed, the moon's cratered aspect can now be explained simply as a result of its high emmental composition, without requiring large amounts of impacts from other hypothetical bodies wandering the solar system. The Moon's formation is similarly explained without the need for an unlikely collision event (we remind the reader that space is very big and thus collisions quite unlikely). Rather, the Moon can have been formed directly in-situ, accumulated from excess cheese escaping the Earth's atmosphere and being captured in orbit.

%--------------------------------------------------------------------

\section{Conclusion}\label{sec5}

In this paper, we have shown evidence for a radically novel theory to explain the Lunar surface's composition: it is made out of cheese. Indeed, our state of the art analysis reveals that a simple model containing 4 common cheese types can effectively and convincingly reproduce the spectrum of various regions on the Moon, thus revealing different cheese contents at separate Lunar locations. We urge the scientific community to seriously consider these novel results and exercise some positive scepticism with regards to previous Lunar composition models. We expect these results to have critical implications for planetary formation and evolution theories, such as a radical shift in our interpretations of the Lunar surface's cratered aspect, which may in fact be due solely to the holes in the cheese. We expect that this revelation will also enable a radical simplification of the Lunar formation theory, which may have just been formed in situ from excess dairy products, rather than requiring an ancient massive planetary impact.

\begin{figure}[h]
\centering
    \includegraphics[scale=0.2]{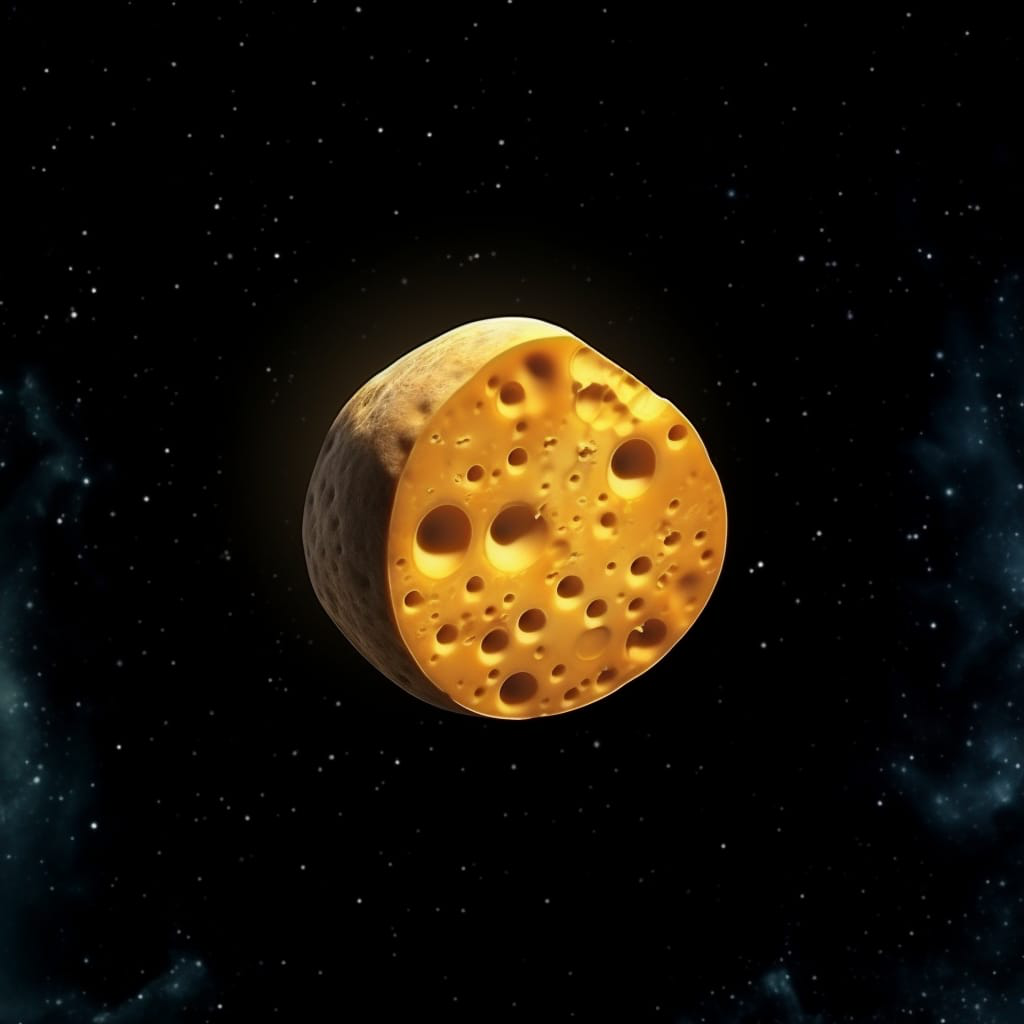}
    \caption{\small{Realistic rendition of what a close-up shot of the cheese Moon would ressemble, generated by the state-of-the-art \textbf{GROMIT} (\textbf{G}enerative \textbf{R}endition \textbf{O}f the \textbf{M}oon's \textbf{I}nferred \textbf{T}exture) simulation.}}\footnotemark
    \label{moon_cheese}
\end{figure}

These groundbreaking findings also have important implications which are not limited to theory. With the advent of large scale space travel, a cheese-filled Moon may become a great strategic asset for space exploration. Indeed, our Moon could then become a greatly important refueling stop to stock up on much needed sustenance for long manned space journeys. In light of this, we suggest that this work should promptly be extended to other moons in our solar system and beyond. Indeed, with recent programs being accepted to try and observe moons outside our solar system using JWST, cheesy exo-moon detections could provide the first clear detection of biosignature outside our solar system, indicative of the presence of complex life forms such as exo-cows!

Moons within our solar system, meanwhile could make for a great network of supply points for space exploration in the near to mid-term future. We suspect that these other Moons would additionally provide very different cheese compositions and thus flavour profiles, and potentially even novel, never before discovered cheese types! The Jovian moons may be a particularly promising system, as Jupiter holds in excess of 80 moons of vastly different shapes and sizes. In particular, we expect the innermost moon, Io, may be one of the only moons to hold melted cheese on its surface, due to the important tidal heating felt by this moon. If this were confirmed, Io could become the ideal location for the first ever astronaut fondue party in outer space!

\footnotetext{i.e. \href{https://www.midjourney.com/home}{Midjourney AI}}

\begin{acknowledgements}

We would like to thank Aardman animation studios for their timeless classic \textit{"Wallace and Gromit"} which has inspired this "research". We hope it will continue inspiring children - young or grown-up like us - around the world into the distant future. We also extend our gratitude to openAI's ChatGPT (a.k.a. \textit{Chapster}, for the well-acquainted) for their invaluable help in writing portions of this manuscript and suggesting acronyms that sometimes even actually worked. We thank the anonymous referee for not giving any comments on - or indeed looking at - this manuscript. We thank Roxane Van den Bossche for providing us with the advanced simulation and rendition of the cheese moon, enabled by midjourney AI. This work has made use of the python \textsc{numpy}, \textsc{scipy}, and \textsc{matplotlib} packages.
\\
Any scientific works which may happen to be cited in this "research" are in no way affiliated with this work, and the silly (luminary?) "results" presented here are not actually meant to discredit their methods or conclusions. \\ 
\\
So long and thanks for all the (April) fish.

\end{acknowledgements}


\section{APPENDIX}

We expect that some remaining sceptics may object to the findings presented here, arguing that the astronauts on board the Apollo missions would have certainly noticed that the lunar surface was made out of cheese. While this may initially sound like a valid point, we argue that it falls out quite naturally that they actually would not have noticed this fact. Indeed, the Moon being devoid of a significant atmosphere and the astronauts thus wearing space suits, it is naturally expected that they could not have smelt the alluring odors of the lunar cheese. As such, it is perfectly logical that they would not have detected this fact, and our theory remains sound. Our advice to these sceptics (and to NASA) is thus unequivocal: next time you send humans to the Moon, don't forget to bring the crackers!

\end{document}